\documentclass[a4paper,11pt,hyper]{JHEP3}
\usepackage{amsfonts,latexsym,graphicx,epsfig,amssymb,amsmath,mathrsfs}

\newcommand{\bm}[1]{\hbox{\boldmath{$#1$}}}
\newcommand{\sbm}[1]{\hbox{\boldmath{\scriptsize$#1$}}}
\newcommand{\dd}{{\rm d}}

\newcommand{\WI}{W^{(2)\,-1}}

\title{Consistency relations and conservation of $\zeta$ in
holographic inflation}
\author{
Jaume Garriga$^{a, b}$, Yuko Urakawa$^c$
\\
a. Departament de F{\'\i}sica Fonamental i Institut de Ci{\`e}ncies del Cosmos, 
Universitat de Barcelona,
Mart{\'\i}\ i Franqu{\`e}s 1, 08028 Barcelona, Spain\\
b. Institute of Cosmology, Department of Physics and Astronomy,\\
Tufts University, Medford, MA 02155, USA\\
c. Department of Physics and Astrophysics, Nagoya University, Chikusa,
Nagoya 464-8602, Japan}

\abstract{It is well known that, in single clock inflation, the curvature perturbation $\zeta$ is 
constant in time on superhorizon scales. In the standard bulk description this
follows quite simply from the local conservation of the energy momentum tensor in the bulk.
 On the other hand, in a holographic description,  the constancy of the curvature perturbation 
must be related to the properties of the RG flow in the boundary theory. Here, we show that, in single clock
holographic inflation, the time independence of correlators of $\zeta$ follows
from the cut-off independence of correlators of the energy
momentum tensor in the boundary theory, and from the so-called
consistency relations for vertex functions with a soft leg.}

\preprint{}

\maketitle

\begin{document}

\section{Introduction}   \label{Sec:Introduction}
In a holographic description of inflation, the renormalization scale
$\mu$ in the boundary theory is expected to correspond to a temporal
coordinate in the bulk. In a recent paper \cite{JYcsv} we investigated the issue 
of time evolution of the curvature perturbation $\zeta$ by considering
a generic deformed CFT at the boundary, in the limit where
conformal perturbation theory is valid along the RG flow between two
nearby fixed points.  We concluded that the two point function for
$\zeta$ is conserved along the RG flow provided that we make the
identification\footnote{This relation may have slow roll corrections,
beyond the leading order in conformal perturbation theory which was
considered in Ref. \cite{JYcsv}.} 
$$
\mu \propto a, 
$$
where $a$ is the cosmological scale factor.

On the other hand, in Ref. \cite{JYcsv} we were not able to show the
conservation of higher order correlators of $\zeta$. This is technically
complicated, due to the presence of semi-local terms in the relation
between higher order cosmological correlators and boundary
correlators. The renormalization of expressions containing such
semi-local terms (where two or more of the points in the correlator
coincide) is not straightforward, and it is hard to check explicitly whether these expressions 
depend on the renormalization scale. To overcome this difficulty, here we
will use a different approach, which does not rely on conformal
perturbation theory. 

Our strategy will be based on the observation that, in a renormalizable
quantum field theory, there is an energy momentum tensor
whose correlators do not depend on the renormalization scale \cite{CCJ}. 
We will also use the so-called consistency relations, which express higher order vertices with a
soft leg in terms of lower order vertices. 
In this way, we can address the conservation of $n$-point correlators of $\zeta$
recursively, starting with the 2-point function.

The paper is organized as follows. In Section \ref{wfp}, we discuss our
setup and conventions. Section \ref{Sec:CR} deals with the consistency
relations involving correlators with soft legs. In Section
\ref{Sec:Holography} we express the correlators of $\zeta$ in terms of
correlators of the energy momentum tensor in the boundary theory. In
Section \ref{Sec:Conservation} we discuss the conditions which are
necessary for the conservation of $\zeta$ from the point of view of the
boundary theory. In Section \ref{Sec:Diffeo} we generalize our arguments
to the case of tensor perturbations. Our conclusions are summarized in
Section \ref{conclusions}.

\section{Wave function prescription}  \label{wfp}
Correlation functions of primordial perturbations can be obtained from the cosmological wave function
\cite{Maldacena02, JYsingle, Nicolis}. In holography, the wave function of long wavelength modes
is related to the generating functional of the boundary quantum field theory (QFT). For the moment, however, 
we will not assume the holographic relation and we
will simply discuss the correlation functions obtained from a given wave function.

\subsection{Wave functional}
We consider a wave function on a particular time slicing $\Sigma_t$ such that the gauge condition $\delta \phi=0$ is satisfied.
With an appropriate choice of spatial coordinates, we express the $d$-dimensional spatial line element as
\begin{align}
 & \dd l^2_d = a^2(t) e^{2 \zeta(t,\, \sbm{x})} \dd \bm{x}^2\,. \label{Exp:dld}
\end{align}
Here, we neglected the tensor perturbation, which will be discussed in Sec.~\ref{Sec:Diffeo}. We assume that the wave function of the $(d+1)$
dimensional bulk spacetime is given by a functional of the curvature perturbation $\zeta$ on the slicing $\delta \phi=0$,
\begin{align}
 & \psi_t = \psi_t [\zeta(t,\, \bm{x})],
\end{align}
which will become a good approximation in case the universe is dominated by a single scalar degree of freedom.

The probability distribution function is given by 
\begin{align}
 P_t[\zeta] = \left| \psi_t[\zeta] \right|^2 \equiv e^{- W_t[\zeta]}
 \,, \label{Exp:PDF}
\end{align}
and satisfies the normalization condition
\begin{align}
 & \int D\zeta \, P_t[\zeta] =1\,. \label{Cond:norm}
\end{align}
The $n$-point functions for $\zeta$ on $\Sigma_t$ can then be obtained as
\begin{align}
 & \langle \zeta(\bm{x}_1) \zeta(\bm{x}_2)  \cdots  \zeta(\bm{x}_n) \rangle = 
  \int D \zeta \, P_t[\zeta]\,
 \zeta(\bm{x}_1) \zeta(\bm{x}_2)  \cdots \zeta(\bm{x}_n) \,. \label{npoint}
\end{align}
In single clock inflation, where only the adiabatic mode is relevant, $\zeta$
becomes time independent on superhorizon scales. In single field models this
happens once the decaying mode becomes negligibly small. 

For later use, we expand $W_t[\zeta] =- \ln P_t[\zeta]$ as
\begin{align}
 & W_t[\zeta] = \sum_{n=2}^\infty \frac{1}{n!} \int \dd^d \bm{x}_1 \cdots \int \dd^d \bm{x}_n\,
 W^{(n)}(t; \bm{x}_1,\,\cdots\,,\bm{x}_n) \zeta(\bm{x}_1) \cdots
 \zeta(\bm{x}_n),\label{taylor}
\end{align}
where we introduced the vertex functions
\begin{align}
 & W^{(n)}(t; \bm{x}_1,\, \cdots,\,\bm{x}_n)\equiv  
 \frac{\delta^n W_t[\zeta]}{\delta\zeta(\bm{x}_1)  \cdots \delta \zeta(\bm{x}_n)} \bigg|_{\zeta=0} \,. \label{Def:Wn}
\end{align} 
As we shall see in Section \ref{Sec:CR}, the tadpole term with $W^{(1)}$ is required to vanish by Diff invariance.

\subsection{Tree level $\zeta$ correlators from the wave function}  \label{SSec:correlator}

Assuming that the amplitude of $\zeta$ is perturbatively small, the $n$-point function of $\zeta$
can be given in terms of $W^{(m)}$ with $m\leq n$. A more detailed
discussion of this perturbative expansion can be found in Ref.~\cite{JYsingle}. For later use, 
here we reproduce the tree level expressions for the lowest order correlators. 

The power spectrum of $\zeta$
is given by using the inverse matrix of $W^{(2)}(\bm{x}_1,\, \bm{x}_2)$ as
\begin{align}
 & \langle \zeta(\bm{x}_1) \zeta(\bm{x}_2) \rangle  =
 \WI (\bm{x}_1,\, \bm{x}_2)\,. \label{Exp:2pzeta}
\end{align}
We assume invariance under global translations and
rotations. Then, we can express $W^{(n)} $ in Fourier space as
\begin{align}
 &(2\pi)^d \delta \left( \sum_{i=1}^n \bm{k}_i \right)
 \hat{W}^{(n)} \left(\bm{k}_1,\,  \cdots ,\, \bm{k}_n \right)  \equiv
   \prod_{i=1}^n \int \dd^d \bm{x}_i\, e^{-i \sbm{k}_i
 \cdot \sbm{x}_i} W^{(n)}(\bm{x}_1,\, \cdots ,\,\bm{x}_n)\,.  \label{Def:hatWn}
\end{align}
Using the Fourier mode $\hat{W}^{(2)}(k)$ with $k\equiv |\bm{k}|$, the
power spectrum of the curvature perturbation is given by
\begin{align}
 & \langle \zeta (\bm{k}_1) \zeta(\bm{k}_2) \rangle =
 (2\pi)^d \delta (\bm{k}_1 + \bm{k}_2) P(k_1) \label{Exp:W2kF}
\end{align}
with 
\begin{align}
 & P(k) = \frac{1}{\hat{W}^{(2)}(k)}\,.  \label{Exp:Pzeta}
\end{align}

The bi-spectrum for $\zeta(\bm{x})$ is expressed by the cubic
interaction $W^{(3)}(\bm{x}_1,\, \bm{x}_2,\, \bm{x}_3)$ as 
\begin{align}
 & \langle \zeta(\bm{x}_1) \zeta(\bm{x}_2) \zeta(\bm{x}_3) \rangle = - \int
 \prod_{i=1}^3  \dd^d \bm{y}_i\, \WI(\bm{x}_i\,,\bm{y}_i)\,
 W^{(3)} (\bm{y}_1,\,\bm{y}_2,\,\bm{y}_3)\,,   \label{Exp:BSzeta}
\end{align} 
Here, we need the minus sign, since the three-point vertex is given by
$- {W}^{(3)}(\bm{x}_1,\, \bm{x}_2,\, \bm{x}_3)$. In Fourier space, we have 
\begin{align}
  \langle \zeta(\bm{k}_1) \zeta(\bm{k}_2) \zeta(\bm{k}_3)
 \rangle_{\rm conn}
 &= (2\pi)^d \delta (\bm{k}_1 + \bm{k}_2 + \bm{k}_3)\,
 B\left( k_1,\, k_2,\, k_3 \right) \,  
\end{align}
with
\begin{align}
 & B\left(k_1,\, k_2,\, k_3 \right) = -  \frac{ \hat{W}^{(3)} \left(k_1,\, k_2,\,
 k_3 \right)}{\hat{W}^{(2)}(k_1) \hat{W}^{(2)}(k_2)
 \hat{W}^{(2)}(k_3)} = - \hat{W}^{(3)} \left(k_1,\, k_2,\,
 k_3 \right) \prod_{i=1}^3 P(k_i)\,. \label{Exp:Bzeta}
\end{align}

The tri-spectrum is composed of the two-different diagrams (see Fig. 2
of Ref.~\cite{JYsingle}). In Fourier space, it is given by
\begin{align}
 \langle \zeta(\bm{k}_1) \zeta(\bm{k}_2) \zeta(\bm{k}_3) \zeta(\bm{k}_4)
 \rangle_{\rm conn} &= (2\pi)^d\, \delta\left( \sum_{i=1}^4 \bm{k}_i\right) \,
  T \left( \bm{k}_1,\,\bm{k}_2,\, \bm{k}_3,\, \bm{k}_4 \right)\,,
\end{align}
with
\begin{align}
 & T (\bm{k}_1,\, \bm{k}_2,\, \bm{k}_3,\, \bm{k}_4) = T_1 \left( \bm{k}_1,\,\bm{k}_2,\, \bm{k}_3,\, \bm{k}_4 \right) +
 T_2 \left( \bm{k}_1,\,\bm{k}_2,\, \bm{k}_3,\, \bm{k}_4 \right) \cr
 & \qquad \qquad \qquad \qquad \qquad \qquad   + T_2
 \left( \bm{k}_1,\,\bm{k}_3,\, \bm{k}_2,\, \bm{k}_4 \right) + T_2 \left(
 \bm{k}_1,\,\bm{k}_4,\, \bm{k}_3,\, \bm{k}_1 \right) \,,  \label{Def:T} \\
 & T_1(\bm{k}_1,\, \bm{k}_2,\, \bm{k}_3,\, \bm{k}_4)  = - 
 \hat{W}^{(4)} \left(\bm{k}_1,\, \bm{k}_2,\,\bm{k}_3,\, \bm{k}_4 \right)
  \prod_{i=1}^4 P(k_i) \,,  \label{Exp:T1} \\
  & T_2(\bm{k}_1,\, \bm{k}_2,\, \bm{k}_3,\, \bm{k}_4)  = \hat{W}^{(3)}\left(
 k_1,\, k_2,\, k_{12} \right) \hat{W}^{(3)}\left(
 k_3,\, k_4,\, k_{34} \right) P (k_{12}) \prod_{i=1}^4 P(k_i)   \,, \label{Exp:T2} 
\end{align}
where we introduced the momentum $\bm{k}_{ij}$ and its absolute value as
$\bm{k}_{ij}\equiv \bm{k}_i + \bm{k}_j$ and $k_{ij}\equiv |\bm{k}_{ij}|$.
Note that using the bi-spectrum $B(k_1,\, k_2,\, k_3)$,
we can express $ T_2(\bm{k}_1,\, \bm{k}_2,\, \bm{k}_3,\, \bm{k}_4)$ as
\begin{align}
 &  T_2(\bm{k}_1,\, \bm{k}_2,\, \bm{k}_3,\, \bm{k}_4)  =  \frac{B \left(
 k_1,\, k_2,\, k_{12} \right)  B \left(
 k_3,\, k_4,\, k_{34} \right)}{P(k_{12})}\,.
 \label{Expo:T2} 
\end{align}
Similarly, we can express the $n$-point function of $\zeta$, using $\hat{W}^{(m)}$
with $m \leq n$.

\section{Consistency relations from diffeomorphism invariance}  \label{Sec:CR}
In this section, we derive a Ward-Takahashi identity from diffeomorphism
invariance (see also Refs.~\cite{Nicolis,Khoury} for a related relevant discussion.) 

Note that the wave function characterizes
the bulk spacetime beyond the tree level perturbative analysis, and can be used in order to
compute correlators to any loop order. 
Here, imposing diffeomorphism invariance on the $\delta \phi=0$ slicing, we derive a
condition on $W^{(n)}$.
When the amplitude of $\zeta$ is sufficiently small, we can perturbatively compute
the correlators of $\zeta$~\cite{JYsingle} from the vertex functions $W^{(n)}$, as discussed in Section \ref{wfp}. 
At the tree level, the condition on $W^{(n)}$ leads to the standard consistency relation for correlators of
$\zeta$. It should be stressed, however, that the condition on
$W^{(n)}$ holds non-perturbatively, and is therefore more fundamental. 
 
\subsection{Dilatation invariance of the wave function}
Among the coordinate transformations, we consider the
dilatation
\begin{align}
 & \bm{x} \to \bm{x}_s \equiv e^s \bm{x}
\end{align}
with a constant parameter $s$, under which the spatial line element,
given in Eq.~(\ref{Exp:dld}), is rewritten as  
\begin{align}
 & \frac{\dd l^2_d}{a^2(t)} = e^{2 \zeta(t,\, \sbm{x})} \dd\bm{x}^2 = e^{2 
 \zeta_s(t,\, \sbm{x}_s)} \dd \bm{x}_s^2 = e^{2 
 \{ \zeta_s(t,\, e^s \sbm{x}) +s \}} \dd \bm{x}^2 \,.
\end{align}
Then, under the dilatation, the curvature perturbation is
changed into
\begin{align}
 & \zeta_s(t,\, \bm{x}) = \zeta(t,\, e^{-s}\bm{x}) -s\,. 
\end{align}
The change of the coordinates
$\Delta \bm{x} \equiv \bm{x}_s - \bm{x}$ 
increases with distance to the origin. However, the change $\Delta \bm{x}$ can stay perturbatively small in the observable region, which is necessarily 
bounded. Similar (residual) gauge transformations which diverge in the limit $\bm{x} \to \infty$
can be found in other gauge theories, such as QED, and it is known that
soft theorems can be derived by using such residual gauge
transformations (see, e.g., Refs.~\cite{Harvey:1996ur,
Avery:2015rga,Mirbabayi:2016xvc}).

The diffeomorphism invariance of the wave function requires that 
the probability distribution function  $P_t[\zeta]=e^{-W_t[\zeta]}$
should be invariant under the dilatation\footnote{Correlation functions
computed by using a Diff invariant probability distribution and measure of
integration are, strictly speaking, ill defined. To make them well defined we must factor out the
infinite volume of the orbits of the gauge group. In this case, $s$ is
the additive parameter in the dilatation group, and so we need to factor
out $\int \dd s$.
This is easily achieved by the standard Fadeev-Popov (FP) trick of introducing a Gaussian factor $\exp\{-(G[\zeta])^2/\alpha\}$ in the integrand, accompanied by the determinant $|\partial G[\zeta]/\partial s|$. Here $\alpha>0$ is an arbitrary positive constant, and $G[\zeta]$ is an $s$ dependent 
function. A convenient choice for $G$ is the average value of the curvature perturbation in the region of our interest, 
$G[\zeta] \equiv \bar\zeta = (\int \dd^3 x\ \zeta)/(\int \dd^3 x)$. Under gauge transformation, we have  $\bar\zeta\to \bar\zeta-s$. In this case, 
the determinant is constant, and there is no need to introduce FP ghosts. 
In summary, the Diff invariant exponent in the distribution function of
the functional integrand, $W= - \ln P$, gets replaced by $\tilde W = W+\bar\zeta^2/\alpha$. 
It is straightforward to check that this modification does not change
the correlation functions, since (in the limit of infinite volume) 
the second term in $\tilde W$ has vanishing functional derivative with
respect to the curvature perturbation.\label{f2}} 
\begin{align}
 & W_t[\zeta(\bm{x})] = W_t[\zeta(e^{-s}\bm{x}) -s]\,.  \label{Cond:Wsc}
\end{align} 
In Sec.~\ref{Sec:Diffeo}, this argument will be briefly extended 
to include the tensor perturbation~\footnote{Here, we impose the diffeomorphism invariance on the $d$-dimensional time
slicing $\Sigma_t$. When we keep only the $d$-dimensional
diffeomorphism invariance, but we break the $(d+1)$-dimensional
diffeomorphism invariance as in Horava-Lifshitz theory, typically there
appears an additional scalar degree of freedom. This case is excluded in
our setup where the wave function $\psi_t$ is expressed only by the
single (scalar) degree of freedom.}, and we will comment on its implications for the boundary theory.

\subsection{Ward-Takahashi identity}
Next, we derive the Ward-Takahashi identity associated with the
dilatation invariance. The curvature perturbation after the scale transformation is
given by 
\begin{align}
 \zeta(e^{-s}\bm{x}) -s &= \sum_{n=0}^\infty \frac{s^n}{n!} \frac{\dd^n}{\dd
 s^n} \zeta(e^{-s} \bm{x}) \Big|_{s=0} - s \cr
  &= \zeta(\bm{x}) - s(\bm{x}
 \cdot \partial_{\sbm{x}} \zeta(t,\, \bm{x}) + 1)   + \sum_{s=2}^\infty
 \frac{(-s)^n}{n!} (\bm{x} \cdot \partial_{\sbm{x}})^n \zeta(\bm{x}) \,, \label{Exp:zetas}
\end{align}
where on the second equality, we replaced $\dd/\dd s$ with 
$- \bm{x} \cdot \partial_{\sbm{x}}$. Using
Eq.~(\ref{Exp:zetas}), we find that at ${\cal O}(s)$,
Eq.~(\ref{Cond:Wsc}) gives
\begin{align}
 & {\cal O}(s): \qquad \quad 0= \int \dd^d \bm{x} \frac{\delta
 W_t[\zeta]}{\delta \zeta(\bm{x})} \{1 + \bm{x} \cdot \partial_{\sbm{x}}
 \zeta(\bm{x})\}\,,  \label{WTsc1}
\end{align}
and at ${\cal O}(s^2)$, it gives
\begin{align}
 & {\cal O}(s^2):  \quad 0=\int \dd^d \bm{x} \frac{\delta
 W_t[\zeta]}{\delta \zeta(\bm{x})} (\bm{x} \cdot \partial_{\sbm{x}})^2
 \zeta(\bm{x}) \cr
 &\qquad \qquad \qquad \quad  + \int d^d \bm{x}_1 \! \int \dd^d \bm{x}_2 \frac{\delta^2
 W_t[\zeta]}{\delta \zeta(\bm{x}_1) \delta \zeta(\bm{x}_2)} \{1 + \bm{x}_1 \cdot \partial_{\sbm{x}_1}
 \zeta(\bm{x}_1)\}\{1 + \bm{x}_2 \cdot \partial_{\sbm{x}_2}
 \zeta(\bm{x}_2)\}\,,  \cr  \label{WTsc2}
\end{align}
and so on. 

It follows from (\ref{WTsc1}) with $\zeta({\bm x})=0$ that $\int \dd^d{\bm x}\, W^{(1)} =0$. By translation invariance, this implies 
\begin{equation}
W^{(1)}= 0. 
\end{equation}
In particular, the tadpole term in (\ref{taylor}) will not contribute, even if the average value of the curvature perturbation $\bar\zeta$ is non-vanishing.
Likewise, from (\ref{WTsc2}) with $\zeta(\bm{x}) =0$, we find $\int
\dd^d{\bm x}_1 \dd^d{\bm x}_2\, W^{(2)} (t;{\bm x}_1,{\bm x}_2) =0$. 
It follows that, in momentum space,
\begin{equation}
\hat W^{(2)}(0,0)=0. 
\end{equation}
This means that the tree level dispersion of $\hat\zeta({\bm k}=0)$ will be infinite\footnote{The FP Gaussian $e^{-\bar\zeta^2/\alpha}$ discussed in footnote \ref{f2}
will make the dispersion of $\bar\zeta$ finite, but the dispersion of $\hat\zeta({\bm k}=0)$ will still be infinite, since both variables are related by an infinite volume factor. Note that 
$\hat\zeta(\bm{k})=\bar\zeta \,\delta(\bm{k}=0)+...$, where the ellipsis
denote contributions with $\bm{k} \neq 0$.}.

\subsection{Consistency relation for vertices with one soft leg}\label{sl}
Taking $(n-1)$-derivatives with respect to
$\zeta(\bm{x}_i)$ for $i=1, \cdots, n-1$ on Eq.~(\ref{WTsc1}), we obtain  
\begin{align}
 & 0=  \int \dd^d \bm{x} \Biggl[ \{1 + \bm{x} \cdot \partial_{\sbm{x}}
 \zeta(\bm{x})\} \frac{\delta^n W_t[\zeta]}{ \delta \zeta(\bm{x}) \delta \zeta(\bm{x}_1)
 \cdots \delta \zeta (\bm{x}_{n-1}) } \cr
 & \qquad \qquad   + \sum_{i=1}^{n-1} \bm{x} \cdot
 \partial_{\sbm{x}} \delta(\bm{x} - \bm{x}_i)  \frac{\delta^{n-1}
 W_t[\zeta]}{ \delta \zeta(\bm{x}) \cdots  \delta \zeta(\bm{x}_{i-1})
 \delta \zeta(\bm{x}_{i+1}) \cdots  \cdots \delta \zeta (\bm{x}_{n-1}) } \Biggr].
\end{align}
Notice that the functional derivative in the second term excludes the
derivative with respect to $\zeta(\bm{x}_i)$. Performing the integration
by parts and setting $\zeta=0$, we obtain
\begin{align}
 &\int \dd^d \bm{x} W^{(n)} (t;\, \bm{x},\, \bm{x}_1,\, \cdots,\,
 \bm{x}_{n-1}) \cr
 & \qquad \qquad \qquad - \sum_{i=1}^{n-1} \partial_{\sbm{x}_i} \left\{ \bm{x}_i W^{(n-1)} (t;\,
 \bm{x}_1,\, \cdots,\,  \bm{x}_{n-1})\right\}  =0 \,. \label{WTsc1n}
\end{align}
This is the generalized consistency relation for $\zeta$, which does not
require the validity of the perturbative analysis. 

It may be convenient to express Eq.~(\ref{WTsc1n}) in Fourier
space. Multiplying Eq.~(\ref{WTsc1n}) by
$
 \prod_{i=1}^n \int \dd^d \bm{x}_i e^{- i \sbm{k}_i \cdot \sbm{x}_i},
$
and using Eq.~(\ref{Def:hatWn}), we express Eq.~(\ref{WTsc1n}) as
\begin{align}
 & 0= \hat{W}^{(n)} (t;
 \bm{k}=0,\, \{ \bm{k}_i \}_{n-1}) + \left(
 \sum_{i=2}^{n-1} \bm{k}_i \cdot \partial_{\sbm{k}_i} -d \right)
 \hat{W}^{(n-1)}\! \left(t; \{ \bm{k}_i \}_{n-1} \right) \,,  \label{Cond:CR1}
\end{align}
where we used
\begin{align}
 & \sum_{i=1}^{n-1} \bm{k}_i \cdot \partial_{\sbm{k}_i} \left[ \delta\! \left( \sum_{i=1}^{n-1} \bm{k}_i
 \right) \hat{W}^{(n-1)} (t;\{ \bm{k}_i \}_n)  \right]  \cr
 &  =   \delta\! \left( \sum_{i=1}^{n-1} \bm{k}_i \right) \times \left(
 \sum_{i=2}^{n-1} \bm{k}_i \cdot \partial_{\sbm{k}_i} -d \right)
 \hat{W}^{(n-1)}\! \left(t; - \bm{K}_{2, n-1},\,\bm{k}_2,\,
 \cdots,\, \bm{k}_{n-1} \right) \,, \label{Exp:ipF}
\end{align}
and removed a delta function which appears as a common factor in the
two terms. Here, $\{ \bm{k}_i \}$ denotes $(n-1)$ momenta $\bm{k}_i$ with $i=1, \cdots,\, n-1$ which satisfies 
$\sum_{i=1}^{n-1} {\bm k}_i=0$ and in the second line, we replaced
$\bm{k}_1$ with $\bm{K}_{m, n}$ defined as
\begin{align}
 & \bm{K}_{m, n} \equiv  \sum_{i=m}^{n} \bm{k}_i\,. 
\end{align}
Equation (\ref{Exp:ipF}) can be verified by operating 
$\int \dd^d \bm{k}_1$ on the both sides. Notice
that Eq.~(\ref{Cond:CR1}) states that if $\hat{W}^{(n-1)}$ does not
depend on time, neither does $\hat{W}^{(n)}$ with one soft leg.

When the amplitude of $\zeta$ is perturbatively small, as discussed in
Appendix \ref{Sec:ACR}, Eq.~(\ref{Cond:CR1}) simply yields the 
consistency relation in a $(d+1)$ dimensional spacetime, given by 
\begin{align}
 & \lim_{k_n \to 0} \frac{{\cal C}^{(n)}(\{\bm{k}_i\}_n)}{P(k_n)} = -
 \left( \sum_{i=2}^{n-1} \bm{k}_i \cdot 
 \partial_{\sbm{k}_i} + d(n-2) \right) {\cal
 C}^{(n-1)}(\{\bm{k}_i\}_{n-1})\,, \label{Exp:CRd}
\end{align}
where ${\cal C}^{(n)}$ denotes the $n$-point function of $\zeta$ with the momentum conservation factor
$$
(2\pi)^d \delta \left( \sum_{i=1}^{n} \bm{k}_i \right)
$$
removed. In Ref.~\cite{Hinterbichler:2013dpa}, the consistency
relation (\ref{Exp:CRd}) was derived for $d=3$. The argument in Appendix
\ref{Sec:ACR} shows that an extension to a general spacetime dimension
proceeds straightforwardly. The consistency relation involves a 
soft mode which is induced by a coordinate transformation. Therefore, we do not expect
any influence of such soft mode in correlators of a variable which remains
invariant under the dilatation. This was explicitly shown in Refs.~\cite{Tanaka:2011aj, Pajer:2013ana}.

In rewriting the WT identity (\ref{Cond:CR1}) in the form
(\ref{Exp:CRd}), we implicitly assume the continuity of $\hat W^{(3)}$
at ${\bm k}=0$. It has been argued in Refs.~\cite{Khoury, Tanaka:2015aza} that this follows from the constancy
of $\zeta$ at long wavelengths (see also Ref.~\cite{Weinberg}). For the purposes of this paper, we will
use the consistency relation in the form (\ref{Cond:CR1}). In the
holographic context, its continuity at ${\bm k=0}$ requires separate
justification. We will come back to this issue in Section
\ref{Sec:Conservation}. 

From Eq.~(\ref{WTsc2}), we can perturbatively derive the consistency relation which
connects the $n$-point function of $\zeta$ with two soft legs to the
$m$-point functions with $m< n$. This also can be derived by sending another momentum to 0 in the consistency
relation (\ref{Cond:CR1}) with one soft leg.

\section{$\zeta$ correlators from holography}    \label{Sec:Holography}
Our previous discussion is based on the use
of the wave function for single field inflationary models. Our next task is to introduce 
the holographic duality between the
$(d+1)$-dimensional cosmological spacetime and the $d$-dimensional
field theory at the boundary. The gauge/gravity duality in the inflationary setup is discussed, e.g., in 
Refs.~\cite{Maldacena02, JYsingle, Strominger, Strominger2, Witten, Bousso:2001mw,
Harlow:2011ke, Anninos:2011ui, LM03, LM04, Seery:2006tq, vdS, Shiu,
Mata:2012bx, Ghosh:2014kba, Banks:2013qra, Banks:2013qpa, Pimentel13,
MS_HC09, MS_HC10, MS_HCob10, MS_NG, MS_NGGW, Kiritsis:2013gia, BMS,
Kundu:2014gxa, KST15, gv, Garriga:2014fda, Garriga:2015tea, McFadden:2013ria, McFadden:2014nta, Kawai:2014vxa}. 

\subsection{Holographic prescription}

Following Ref.~\cite{HS}, we will assume that the probability
distribution of the bulk gravitational field is related to the
generating functional of a boundary QFT as
\begin{align}
 & P[h,\, \phi]= |\psi_{\rm bulk}[h,\phi]|^2 \propto Z_{\rm QFT}[h,\phi] Z^*_{\rm QFT}[h,\phi]
 \,, \label{duality}
\end{align}
where the generating functional $Z_{\rm QFT}$ is given by
\begin{equation}
Z_{\rm QFT} [h,\phi]  =  e^{-W_{\rm QFT}[h,\phi]} =  \int D \chi\,
\exp \left(  - S_{\rm QFT} [\chi,\, h,\, \phi]\right)   \, \label{Exp:Z}.
\end{equation}
Here, $\chi$ stands for the set of boundary fields. In the boundary, the
path integral is doubled by multiplying the generating functional and
its complex conjugate together. Comparing Eq.~(\ref{duality}) to Eq.~(\ref{Exp:PDF}), we find that
$W^{(n)}$ are given by
\begin{align}
 & W^{(n)}(t(\mu); \bm{x}_1,\, \cdots,\,\bm{x}_n) = 2 {\rm Re} \left[ 
 \frac{\delta^n W_{\rm QFT}[\zeta]}{\delta\zeta(\bm{x}_1)  \cdots \delta
 \zeta(\bm{x}_n)} \bigg|_{\zeta=0}  \right] \,. \label{Def:WnH}
\end{align}  
Here, we expressed the time dependence in terms of the renormalization scale $\mu$,
postulating that the time evolution of the bulk spacetime is described by the
renormalization group flow.

\subsection{Vertex functions}

 We may now express $W^{(n)}$ in terms of boundary
correlators of the energy-momentum tensor,
defined by
\begin{align}
 & T_{ij} \equiv - \frac{2}{\sqrt{h}} \frac{\delta S_{\rm QFT}}{\delta h^{ij}}. \label{tij}
\end{align}
The derivative of the boundary action with respect to $\zeta$ is then given by
the trace part of the energy-momentum tensor, as
\begin{align}
 &  \frac{\delta S_{\rm QFT}}{\delta \zeta(\bm{x})} = e^{(d-2)\zeta(\sbm{x})}
 \delta^{ij} T_{ij}[\zeta](\bm{x})\,. \label{Exp:dSdzeta}
\end{align}
Using Eq.~(\ref{Exp:dSdzeta}), we easily find
\begin{align}
  W^{(2)}(\bm{x}_1,\, \bm{x}_2)&= - 2 {\rm Re} \left[ \left\langle \frac{\delta S_{\rm QFT}}{\delta \zeta(\bm{x}_1)}
 \frac{\delta S_{\rm QFT}}{\delta \zeta(\bm{x}_2)} \right\rangle \bigg|_{\zeta=0}
 \right]= - 2
 {\rm Re} \left[
 \left\langle T(\bm{x}_1) T(\bm{x}_2) \right\rangle  \right]\,, \label{Exp:W2}  
\end{align}
and
\begin{align}
  W^{(3)}(\bm{x}_1,\, \bm{x}_2,\, \bm{x}_3)
 &= 2{\rm Re} \left[  \left\langle \frac{\delta S_{\rm QFT}}{\delta \zeta(\bm{x}_1)}
 \frac{\delta S_{\rm QFT}}{\delta \zeta(\bm{x}_2)}  \frac{\delta S_{\rm QFT}}{\delta \zeta(\bm{x}_3)}
   - \left\{  \frac{\delta S_{\rm QFT}}{\delta \zeta(\bm{x}_1)}
 \frac{\delta^2 S_{\rm QFT}}{\delta \zeta( \bm{x}_2) \delta \zeta(\bm{x}_3)}   + (2\, {\rm perms})  \right\}
 \right\rangle \bigg|_{\zeta=0} \right] \cr
 &= 2{\rm Re} \biggl[   \Big\langle T(\bm{x}_1) T(\bm{x}_2) T(\bm{x}_3)    - (d-2) \{ 
  T(\bm{x}_1) T(\bm{x}_2) \delta_\mu(\bm{x}_{23}) +
 \left( 2 {\rm perms}\right) \}  \nonumber \\
 & \qquad \quad \qquad -
 \left\{ T(\bm{x}_1) \partial T(\bm{x}_2) \delta_\mu(\bm{x}_{23})  +  \left(
 2 {\rm perms}\right) \right\} \Big\rangle  \biggr] \,.     \label{Exp:W3}
\end{align}
Similarly, we can calculate the higher order vertex functions $W^{(n)}$ in terms of correlators of the energy momentum tensor.
Here, we introduced
\begin{align}
 & T(\bm{x}) \equiv \delta^{ij} T_{ij}(\bm{x}) |_{\zeta=0},\, \qquad 
  \partial T(\bm{x}) \equiv \frac{\partial T[\zeta](\bm{x})}{\partial \zeta(\bm{x})}
 \bigg|_{\zeta=0}\,. 
\end{align}
The coincidence limit is described by the smeared delta function
$\delta_\mu(\bm{x})$ which takes a non-vanishing value only at
$|\bm{x}| \leq 1/\mu$ and is normalized as
\begin{align}
 & \int \dd^d \bm{x} \delta_\mu (\bm{x}) =1\,. 
\end{align}

Since the right hand side of Eq.~(\ref{Exp:dSdzeta}) does depend on $\zeta$, the
$n$-th derivative with $n\geq 2$ of $S_{\rm QFT}$ does not vanish. Because of that, $W^{(n)}$ with $n\geq 3$ includes semi-local
terms, where some of the arguments $\bm{x}_i$ with $i= 1,\, \cdots,\, n$
coincide, but the rest do not. 
The UV divergence from an ultra-local
term, where all the arguments coincide, can be renormalized by using local counter terms. On the other hand, the regularization
of the semi-local terms is not straightforward (see, e.g., Refs.~\cite{MS_NG,BMS, AK}).
Fortunately, for present purposes we will be able to sidestep this difficulty by using the consistency relations,
as we shall see in the next Section.

Instead of using $\zeta$, one may wish to introduce another
variable $X(\bm{x})$, whose derivatives of $W_t$ do not yield any semi-local
terms~\cite{Ghosh:2014kba,Kundu:2014gxa, KST15}. For that, the boundary
action $S_{\rm QFT}$ should depend on the new variable $X(\bm{x})$ only
linearly. i.e., 
\begin{align}
 & \frac{\delta^2 S_{\rm QFT}}{\delta X(\bm{x}_1) \cdots \delta
 X(\bm{x}_n)} = 0 \qquad \quad (n \geq 2)\,.  \label{Exp:delSX}
\end{align} 
However, if $X$ is a degree of freedom in the metric,
it seems hard to find a variable that satisfies Eq.~(\ref{Exp:delSX}), because the action $S_{\rm QFT}$ non-linearly
depends on the metric~\footnote{If the boundary
action is given by a single trace operator $O$ as 
$$
S_{\rm QFT} = S_{\rm CFT} +  \int \dd \Omega \phi O\,,
$$
we can choose $X(\bm{x})$ as $X(\bm{x}) = \delta \phi(\bm{x})$ (in the
flat gauge) or a variable which is linearly related to $\delta \phi(\bm{x})$ such as
$\zeta_n(\bm{x}) \equiv - (H/\dot{\phi}) \delta \phi(\bm{x}) $, which was
introduced in Ref.~\cite{Maldacena02}. Notice that since $\zeta$ and
$\zeta_n$ are non-linearly related as presented in Eq.~(A.8) of
Ref.~\cite{Maldacena02}, $W^{(n)}$ with $n\geq 3$ for $\zeta$ include
the semi-local terms as we discussed here. 
}.

\section{Conservation of $\zeta$ and consistency relation}  \label{Sec:Conservation}
In this section, we show that the WT identity for dilatation restricts the
correlators of the energy-momentum tensor in the limit where some of the arguments coincide. 
 
\subsection{Cutoff independence of the energy momentum tensor}

In Ref.~\cite{CCJ}, Callan, Coleman, and Jackiw (CCJ) considered the cutoff dependence of the so-called 
improved energy momentum tensor $\Theta_{\mu \nu}$. This differs from the conventional flat space symmetric 
energy momentum tensor by terms which are conserved identically\footnote{The improved energy momentum tensor
can be obtained from the action of matter in curved space, with suitable non-minimal couplings to the
metric, by taking functional derivative with respect to the metric and subsequently taking the flat space limit.
It was shown in Ref.~\cite{Collins} that in order to establish the cut-off independence of the improved 
energy momentum tensor it is important to
consider the running of the non-minimal coupling. More recently, this issue has been
discussed in more detail, e.g., in Refs.~\cite{Yonekura, Nakayama}, following
Ref.~\cite{Polchinski}. For our purposes, it will be sufficient to assume that the cut-off independent 
energy momentum tensor $\Theta_{ij}$, can be obtained from the boundary theory by functional derivative
with respect to the boundary metric, as in Eq.~(\ref{tij}).}. 
CCJ showed that in a renormalizable theory, an insertion of $\Theta_{\mu \nu}$ to the
correlators of the matter fields $\chi$ does not yield any cutoff
dependence. Iterating the argument, it follows that $n$-successive 
insertions of $\Theta_{ij}$ do not give rise to any $\mu$ dependent
contributions, as long as all the points are separated in
position space. When some of the $n$-points coincide, the
WT identity which was used in the discussion of CCJ potentially includes a
momentum integral, which also integrates the UV modes and may induce a
cutoff dependence.

In the following, we choose our energy momentum tensor for the boundary theory
to be the improved energy-momentum tensor, {\it i.e.}, 
$T_{ij}= \Theta_{ij}$. In that case, we can express the correlators of $T_{ij}$ (in the flat space limit) as 
\begin{align}
 & {\rm Re} \left[ \langle T_{ij}(\bm{x}_1) T_{kl}(\bm{x}_2) \rangle
 \right] = {\rm Re} \left[ \langle T_{ij}(\bm{x}_1) T_{kl}(\bm{x}_2) \rangle_0
 \right]\,, \label{Exp:T2} \\
 & {\rm Re} \left[ \langle T_{ij}(\bm{x}_1) T_{kl}(\bm{x}_2) T_{mn}(\bm{x}_3) \rangle
 \right] = {\rm Re} \left[ \langle T_{ij}(\bm{x}_1) T_{kl}(\bm{x}_2) T_{mn}(\bm{x}_3)
 \rangle_0 \right] \cr
 & \qquad \qquad \qquad \qquad \qquad \qquad \qquad \qquad   + \delta_\mu (\bm{x}_{12}) {\cal F}_{ij;kl;mn}(\mu;
 \bm{x}_{23})  + (2 \, {\rm perms})\,, \label{Exp:T3}
\end{align} 
where $\langle T_{ij}(\bm{x}_1) T_{kl}(\bm{x}_2) \rangle_0$ and $\langle T_{ij}(\bm{x}_1) T_{kl}(\bm{x}_2) T_{mn}(\bm{x}_3) \rangle_0$ denote
the $\mu$ independent contributions of the improved energy momentum
tensor. CCJ's argument does not exclude the appearance
of $\mu$ dependent contributions in the coincidence limit when two
points coincide. The function ${\cal F}_{ij;kl;mn}(\mu;
\bm{x}_{12})$ denotes the possible $\mu$ dependence from the coincidence
limit. In Eqs.~(\ref{Exp:T2}) and (\ref{Exp:T3}), we
dropped the ultra-local terms, since these can be canceled by the local 
counterterms. Thus, the $\mu$ dependence can appear only in $n$-point
correlators with $n>2$.

In a bulk description, the local divergences correspond to a
rapidly oscillating phase of the wave function $\Psi$, which cancels out
in $|\Psi|^2$. In this paper, following Ref.~\cite{HS}, we will adopt
the prescription (\ref{duality}), where the divergent phase
contributions are canceled. (The phase contribution was briefly
discussed in Ref.~\cite{Maldacena02} and in more detail in
Ref.~\cite{Harlow:2011ke}.)

\subsection{Consistency relation and coincidence limit}  \label{SSec:CL}
As shown in the previous section, the vertex functions $W^{(n)}$ are
expressed in terms of the correlators of $T$ and its derivative 
with respect to $\zeta$. Using Eqs.~(\ref{Exp:W2}) and (\ref{Exp:T2}),
we obtain the Fourier mode of $W^{(2)}$ as
\begin{align}
 & \delta(\bm{k}_1 + \bm{k}_2) \hat{W}^{(2)}(k_1) = -2 \prod_{i=1,2}
 \int \dd^d \bm{x}_i e^{i \sbm{k}_i \cdot \sbm{x}_i} {\rm Re} \left[
 \langle T(\bm{x}_1) T(\bm{x}_2) \rangle_0  \right] \,.
\end{align}
We find that the CCJ's argument directly implies that $\hat{W}^{(2)}(k)$
is $\mu$ independent or equivalently time independent in the bulk. When
the perturbative expansion is possible, the $\mu$ independence of $\hat{W}^{(2)}(k)$
immediately leads to the $\mu$ independence of the power spectrum of
$\zeta$. In Ref.~\cite{JYcsv}, assuming that the conformal symmetry is
slightly broken by the deformation operator $\int \dd^d \Omega\, g {\cal O}$ in the boundary and solving
the induced RG flow, the conservation of the power spectrum was
explicitly shown. Here, we see that this result is much more general, and follows from the cut-off independence 
of the correlators of the energy-momentum tensor in a generic renormalizable boundary theory.

As we discussed in Sec.~\ref{Sec:CR}, the consistency relation
(\ref{Cond:CR1}) involves $W^{(n)}$ and $W^{(n-1)}$. 
Since $\hat{W}^{(2)}(k)$ is $\mu$ independent, Eq.~(\ref{Cond:CR1}) requires that
$\hat{W}^{(3)}(0,\, k,\, k)$ should be also $\mu$ independent. Among the
terms in $\hat{W}^{(3)}(0,\, k,\, k)$, a possible $\mu$ dependence can
appear only from the terms with
\begin{align}
 & {\cal F}(\bm{x}) \equiv \delta^{ij} \delta^{kl} \delta^{mn}
 {\cal F}_{ij;kl;mn}(\bm{x})
\end{align}
and $\langle T \partial T \rangle$. Now, we find that the $\mu$ independence of 
$\hat{W}^{(3)}(0,\, k,\, k)$ requires
\footnote{What we directly obtain from the $\mu$ independence of
$\hat{W}^{(3)}(0,\, k,\, k)$ is 
$$
 \frac{\partial}{\partial \mu} \left[ \Delta(\mu,\, k=0) + 2
 \Delta(\mu,\, k)  \right] = 0\,.
$$
First, we set $k=0$,
then we find that $\Delta(\mu,\, 0)$ is $\mu$ independent, which implies 
$\Delta(\mu,\, k)$ with $k \neq 0$ is also $\mu$ independent.} 
\begin{align}
 &  \frac{\partial}{\partial \mu} \Delta(\mu,\, k) = 0 \,, \label{Rel:FpT}
\end{align}
where we defined
\begin{align}
 & \Delta(\mu,\, k) \equiv \hat{\cal F}(\mu;\, k) - {\rm Re} \left[
\langle T \partial T \rangle \right] (\mu;\ k)\,. 
\end{align}
The condition (\ref{Rel:FpT})
implies that the semi-local term $\langle T(\bm{x})\partial T(\bm{y}) \rangle$ 
should be related to the three-point function of $T$ in
the limit where two of the three arguments coincide,
$\lim_{\sbm{z} \to \sbm{x} }\langle T(\bm{x})T(\bm{y})T(\bm{z}) \rangle$.
The possible singular contribution which may arise in
such coincidence limit should cancel out in the combination: 
$$
 {\rm Re} \left[ \lim_{\sbm{z} \to \sbm{x} }\langle
 T(\bm{x})T(\bm{y})T(\bm{z}) \rangle - \langle \partial T(\bm{x})
 T(\bm{y}) \rangle \right]. 
$$
It may seem surprising that a non-trivial condition is obtained from the
requirement of the Diff invariance. This is because we are using
$\zeta$ as our variable in the wave function, and this transforms under
dilatation.
If we could use a Diff invariant variable, we would not obtain any
additional condition. In the bulk description, a technical difficulty to
use such a Diff invariant variable for canonical quantization was
pointed out in Refs.~\cite{Urakawa:2010it, Urakawa:2010kr, review}.

\subsection{Conservation of $\zeta$} \label{coz}

In the previous subsection, we showed that CCJ's argument and
the consistency relation imply the $\mu$ independence of 
$\hat{W}^{(3)}(0,\,k,\, k)$, which leads to the condition (\ref{Rel:FpT}). Since $\Delta(k) \equiv \Delta(\mu,\, k)$ is $\mu$
independent, we may write
\begin{align}
  \hat{W}^{(3)}(k_1,\, k_2,\, k_3)
 &= 2 {\rm Re} [ \langle T T T \rangle_0 ](k_1,\, k_2,\, k_3)   \cr
 & \qquad \qquad  + 2 \sum_{i=1}^3 \left\{ \Delta(k_i) - (d-2)
 {\rm Re} [ \langle TT \rangle_0](k_i)  \right\}\,, 
\end{align}
where ${\rm Re} [\langle TT \rangle_0](k)$ and ${\rm Re} [ \langle T T T \rangle_0 ](k_1,\, k_2,\, k_3)$
denote the $\mu$ independent contributions in the Fourier modes of the
real parts of the two and three point functions for $T$. In this way, 
the $\mu$ independence of  $\hat{W}^{(3)}(k_1,\,k_2,\, k_{12})$ is established.

The $\mu$ independence of  $\hat{W}^{(n)}$ for $n>3$ can be derived recursively by a similar argument,
starting with the assumption that $W^{(n-1)}$ is $\mu$ independent. 
The vertex functions $W^{(n)}$ are given by
the correlators of $T$ and 
$\delta^m S_{\rm QFT}/\delta \zeta(\bm{x}_1) \cdots \delta \zeta(\bm{x}_m) |_{\zeta=0}$ with 
$m < n$. The correlators with $\delta^m S_{\rm QFT}/\delta
\zeta(\bm{x}_1) \cdots \delta \zeta(\bm{x}_m) |_{\zeta=0}$ contribute to
the coincidence limit where $\bm{x}_1,\, \cdots,\,\bm{x}_m$ agree, and
they can depend on $\mu$.  Meanwhile, in this coincidence limit, the
auto-correlation functions of $T$ also can depend on $\mu$. 
The consistency relation then requires that $W^{(n)}$ with one or more of the momenta set to $0$ should be $\mu$ independent.
This determines a relation between all $\mu$ dependent contributions, so that they cancel out in the vertex with a soft leg. From that relation, 
it can be shown that the vertex function $W^{(n)}$ without any soft leg is also $\mu$ independent. Finally, the $\mu$ independence of all $W^{(n)}$ ensures that the probability 
distribution $P[\zeta] = e^{-W[\zeta]}$  (and therefore all  
correlators of $\zeta$) are ``time'' independent, once we interpret the
RG flow as time evolution.

We noted in Subsection \ref{sl} that, in order to rewrite the consistency relation in terms of $\zeta$ correlators at the tree level, in the form Eq.~(\ref{Exp:CRd}), it is necessary to assume 
the continuity of the vertex function $W^{(n)}$ at $\bm{k}=0$. In the holographic context, this 
is equivalent to the continuity of the energy momentum tensor (and its derivative) at $\bm{k}=0$. 

Finally, let us note that in Ref.~\cite{CCJ} the infrared behaviour of correlators with an insertion of $T_{ij}$ was assumed to be analytic. 
In this situation, the vertex functions of the boundary theory are also expected to be analytic in the infrared, and the $\mu$
independence of $W^{(n)}(\{ \bm{k}_i\}_{n})$ follows directly from the $\mu$ independence of 
$W^{(n)}(\bm{k}=0,\, \{\bm{k}_i\}_{n-1})$. In that case the correction for a finite $\bm{k}$ will be suppressed by $(k/\mu)^p$ with an integer positive power
of $p$, and will vanish in the long wavelength limit. Notice that, in the present holographic context, where the CCJ
argument can be applied, a possibility that
$\zeta$ has a growing mode even in one field models (see, e.g.,~Ref.~\cite{Namjoo:2012aa}) is excluded.

\subsection{Primordial spectra from holographic inflation}
When the amplitude of $\zeta$ is sufficiently small, we can compute the
spectra of $\zeta$ using the formulae derived in Section \ref{SSec:correlator}. 
Inserting $\hat{W}^{(2)}$ and $\hat{W}^{(3)}$ into
Eqs.~(\ref{Exp:Pzeta}) and (\ref{Exp:Bzeta}), we obtain the power 
spectrum and the bi-spectrum for $\zeta$ as
\begin{align}
 & P(k) = - \frac{1}{2 {\rm Re} [ \langle T T \rangle_0 ] (k)} \,, \\
 & B(k_1,\, k_2,, k_3) = \frac{1}{4} \frac{1}{\prod_{i=1}^3 {\rm Re} [
 \langle T T \rangle_0 ] (k_i)} \biggl[ {\rm Re} [ \langle T T T
 \rangle_0 ](k_1,\, k_2,\, k_3)  \cr
 & \qquad \qquad \qquad  \qquad \qquad \qquad - \sum_{i=1}^3
 \left\{ (d-2) {\rm Re} [\langle TT \rangle_0](k_i) - \Delta (k_i)  \right\} \biggr]\,, 
\end{align}
where $k_i$ should satisfy $\sum_{i=1}^3 \bm{k}_i =0$. This formulae
should hold generically for a holographic model which is dual to a
single clock inflation, as far as the diffeomorphism invariance in the
boundary is preserved and the boundary theory is renormalizable. To
compute the bi-spectrum of $\zeta$, we need to know the correlators
in the coincidence limit, described by $\Delta(k)$.

\subsection{Relation to previous work}
In Ref.~\cite{JYcsv}, assuming that the boundary theory is given by a
single trace operator ${\cal O}$ as
\begin{align}
 & S_{\rm QFT} = S_{\rm CFT} + \int \dd^d\Omega \,\phi {\cal
 O}\,. \label{Exp:Ssto} 
\end{align}
we addressed the conservation of $\zeta$. The second term describes the
deformation from the CFT, which is needed to be dual to an inflationary
spacetime. The (dimensionless) coupling constant $\phi$ plays the role
of the inflaton.  By solving the RG flow, it was shown that when 
all the arguments $\bm{x}_i$ with $i=1, \cdots,\, n$ are separated by a distance larger than
$1/\mu$, the correlators of ${\cal O}$ satisfy
\begin{align}
 & Z^{-n/2}(\mu) \langle {\cal O}(\bm{x}_1) \cdots {\cal O}(\bm{x}_n) \rangle_\mu =
 Z^{-n/2}(\mu_0) \langle {\cal O}(\bm{x}_1)
 \cdots {\cal O}(\bm{x}_n) \rangle_{\mu_0}\,, \label{ASM:ren}
\end{align}
where $Z(\mu)$ denotes the wave function renormalization. It follows very simply from this 
relation that $\hat{W}^{(2)}$ is $\mu$ independent.

In Ref.~\cite{JYcsv}, it was assumed that the $n$-point function
of ${\cal O}$ for an arbitrary configuration (including the coincidence
limit, where some of the arguments are closer than $1/\mu$) is given
by Eq.~(\ref{ASM:ren}). Under this assumption, it was shown that $\hat{W}^{(3)}$ cannot be $\mu$ independent,
except for a particular case where the beta function $\beta \equiv \dd \phi/\dd \ln \mu$ 
is given by a simple power law scaling as $\beta \propto \mu^{\lambda}$
with a constant parameter $\lambda$. Thus, the conservation of $W^{(n)}$ for $n \geq 3$ in a more generic RG
flow is not compatible with this assumption. On the other hand this assumption was not particularly well motivated, 
since there is no reason to expect that the expression (\ref{ASM:ren})
can be extrapolated to the coincidence limit. 
As we have shown in this section,
the $\mu$ dependent contribution should exist in the coincidence limit, in order to cancel the 
$\mu$ dependent semi-local contribution,  so that the consistency 
relation (\ref{Cond:CR1}) can be satisfied.

Another possibly relevant issue is that 
the setup given by Eq.~(\ref{Exp:Ssto}) might be too simplistic. In Refs.~\cite{HP10,
FLR10}, a Wilsonian treatment of the holographic RG flow was
explored in the context of AdS/CFT. According to their bulk computations, integrating out the UV
contributions generates multi trace operators in the boundary
theory (the RG flow with multi trace
operators in dS/CFT setup was discussed in Ref.~\cite{Das}). In Refs.~\cite{Lee12,
Lee13}, starting from a boundary theory with multi trace operators, it
was shown that the Einstein gravity emerges after a particular mapping
between the boundary and the bulk quantities. Motivated by these
studies, one may generalize the boundary QFT to a more general theory
which includes both single and multi trace 
operators as
\begin{align}
 & S_{\rm QFT}[\phi,\, {\cal O}] = S_{\rm CFT} + \int \dd^d\Omega \phi {\cal O} +  \int \dd^d \Omega {\cal
 F}[\phi] {\cal O} \partial {\cal O} + \cdots\,, 
\end{align}     
where $\partial$ is a derivative operator, and the ellipsis denote 
possible multi trace operators with more ${\cal O}$. The coefficients $\phi$ and ${\cal F}[\phi]$ will depend on 
the bulk gravity theory. It might be interesting to see if the consistency
relation (\ref{Cond:CR1}) imposes any meaningful restriction on the multi-trace operators or
not. We leave this for a future study.

\section{Condition on boundary theory from diffeomorphism invariance}
\label{Sec:Diffeo} 
So far, we have considered only the WT identity for the dilatation,
which restricts the vertex function for the scalar perturbation and, in
turn, the trace part of the energy-momentum tensor in the boundary. As
was pointed out in Refs.~\cite{Urakawa:2010it, Urakawa:2010kr} (see also
Ref.~\cite{Hinterbichler:2012nm}), the gauge: 
\begin{align}
 & \delta \phi=0,\, 
\end{align}
and 
\begin{align}
 & \dd l^2_d = h_{ij} \dd x^i \dd x^j = a^2 e^{2 \zeta} [e^{\gamma}]_{ij} \dd x^i \dd x^j
\end{align}
with
\begin{align}
 & \partial_i \gamma_{ij} = \gamma_{ii} = 0\,,
\end{align}
accommodates an infinite number of residual gauge degrees of freedom (in
the choice of the spatial coordinates), when we are concerned only with
a causally connected region to us. These residual gauge degrees of freedom describe the variation of the boundary
conditions imposed at the null boundary of the causally connected
region~\footnote{It was shown that these residual gauge degrees of freedom give
rise to the infrared divergence of $\zeta$ and $\gamma_{ij}$ (see, e.g.,
Refs.~\cite{review, Tanaka:2012wi, Tanaka:2013xe, Tanaka:2014ina}).}.

Among these residual gauge degrees of freedom, we consider the
transformation 
\begin{align}
 & x^i\, \to \, \tilde{x}^i \equiv \Lambda^i\!_j x^j  \equiv
 (\delta^i\!_j + \delta \Lambda^i\!_j  ) x^j
\end{align}
with   
\begin{align}
 & \Lambda^i\!_j \equiv e^s [e^{S}]^i\!_j\,,
\end{align}
which additionally includes a constant symmetric tensor $S_{ij}$. Under this transformation, the
above gauge condition remains unchanged. The diffeomorphism invariance
of the wave function $\psi_t$ requires that $W_t$ should remain unchanged under this
residual gauge transformation as
\begin{align}
 & W_t \left[h_{ij}(\bm{x}) \right] = W_t \left[ \tilde{h}_{ij}(\bm{x})
 \right]\,, \label{WTwithT}
\end{align}
where $\tilde{h}_{ij}$ is related to $h_{ij}$ as 
\begin{align}
 & \tilde{h}_{ij}(\tilde{\bm{x}}) = (\Lambda^{-1})_i\!^k
 (\Lambda^{-1})_j\!^l \,  h_{kl} (\bm{x})\,.
\end{align}
Here, $\Lambda^{-1}$ denotes the inverse matrix of $\Lambda$. Repeating
a similar argument,  Eq.~(\ref{WTwithT}) gives the WT identity
\begin{align}
 & 0= \delta \Lambda^k\!_l \int \dd^d \bm{x} \left[ \frac{\delta
 W_t}{\delta h_{ij}}\, x^l \partial_k h_{ij} +  2 \frac{\delta
 W_t}{\delta h_{il}} h_{ki} \right]\,. \label{WTwithT2}
\end{align}
This is a generalization of Eq.~(\ref{WTsc1}).

As one may expect, the WT identity (\ref{WTwithT2}) is nothing but the
conservation of the energy momentum tensor on the boundary. In fact, we
can show that Eq.~(\ref{WTwithT2}) is equivalent to 
\begin{align}
 & 0 = \int \dd^d \bm{x} \sqrt{h}\, h_{il} \delta x^l\, \nabla_j {\rm Re}[ \langle T^{ij} \rangle]
\end{align}
with $\delta x^i \equiv \tilde{x}^i - x^i$. Here, we used
\begin{align}
 & \frac{\delta W_t}{\delta h_{ij}} = 2 {\rm Re} \left[ \frac{\delta
 W_{\rm QFT} }{\delta h_{ij}}  \right] = -  \sqrt{h} {\rm Re} \left[
 \langle T^{ij} \rangle \right]\,. 
\end{align}
Since this is derived from the diffeomorphism invariance of
$W_t$, we obtain only the real part of the conservation of the energy
momentum tensor. Meanwhile, the diffeomorphism invariance of 
$W_{\rm QFT}$ gives $\nabla_i \langle T^{ij} \rangle=0$ (see also Ref.~\cite{McFadden:2014nta}). In the gauge
with $\delta \phi=0$, because of the absence of the external source
other than the metric, the expectation value of the energy-momentum
tensor is conserved unlike in the Feffermam and Graham gauge, where the
energy-momentum tensor is not conserved due to the presence of the
external source scalar field~\cite{deHaro:2000vlm}.

Repeating a similar argument to the one in the previous section, with
the use of the WT identity (\ref{WTwithT2}), we can derive the consistency
relation for the scalar and tensor perturbations. (Notice that the
Diff invariance plays the crucial role both in the consistency relation
and the CCJ.) Similarly, the
consistency relation gives the conditions on the coincidence limit, e.g.,
the trace part and the transverse and traceless part of
\begin{align}
 & {\rm Re} \left[ \lim_{\sbm{z} \to \sbm{x} }\langle
 T_{ij}(\bm{x})T_{kl}(\bm{y})T_{mn}(\bm{z}) \rangle - 2 \left\langle
 \frac{\partial T_{ij}(\bm{x})}{\partial h_{mn}(\bm{x})} T_{kl}(\bm{y})
 \right\rangle \right]  \label{semilocal}
\end{align}
should not give rise to singular contributions.

\section{Conclusions} \label{conclusions}

The wave function provides a characterization of the distribution
function $P_t[\zeta]=|\psi_t[\zeta]|^2 \propto e^{ - W_t[\zeta]}$
of primordial fluctuations in the curvature perturbation $\zeta$. 
This characterization does not require the amplitudes of the fluctuations to be perturbatively small, 
and can be applied non-perturbatively. 
In the first half of this paper, we derived the consistency relation for the vertex functions
$W^{(n)}$, which are the coefficients of the expansion of $W_t$ in powers of $\zeta({\bm x})$.
The consistency relation is derived by using the Ward-Takahashi identity associated with dilatation invariance of the wave function, and
provides a generalization of the well-known consistency relation for the tree level correlation functions of $\zeta$,
which relates the $n$-point function with one soft leg to the $(n-1)$-point function.

In the second part of the paper, postulating the holographic duality between
the $(d+1)$-dimensional cosmological spacetime and the $d$-dimensional
boundary theory, we discussed the conservation of the curvature perturbation. 
The main ingredients we used in our discussion are the cut-off independence of correlators
of the energy momentum tensor and the consistency relations amongst vertex functions.
The first assumption is known to be valid in renormalizable field theories, as long as we 
use the appropriately ``improved'' energy momentum tensor. The second ingredient is simply a
consequence of Diff invariance. First, we showed that
the consistency relation provides a condition on the semi-local terms,
where some of the arguments coincide. When this condition is fulfilled, we can show the cutoff independence of
the vertex function $W^{(n)}$. In the bulk perspective, this implies the
conservation of $\zeta$ in time at large scales.

 
An intriguing feature of the holographic approach is that the
conservation of $W^{(2)}$, and hence of the tree level power spectrum
for $\zeta$, follows simply from the cut-off independence of the
correlator of $T$. In a bulk
description, the long wavelength solution for $\zeta$ can have a growing
mode even in one field models, when these exhibit non-attractor
behaviour. This possibility seems to be excluded in the present
holographic context, where the CCJ argument applies. A possible way
around that is to consider a boundary theory where the analyticity in
the IR is violated. In that case we may expect $\mu$ dependence of correlators of
$\zeta$. Conversely, from a bulk perspective, 
non-analyticity in the IR is expected in the case when the long wavelength 
$\zeta$ is not constant in time. The reason is that the shift vector $N_i$ is related to 
$\zeta$ as $\partial^i N_i \propto \dot{\zeta}$, which introduces a non-local
contribution with a negative power of $k$ when we eliminate $N_i$ by
using this relation.

\acknowledgments
We are grateful to David Mateos, Yu Nakayama, Tadakatsu Sakai,
Sergey Sibiryakov, and Takahiro Tanaka for their valuable
comments. Y.~U. would like to thank Institute for Advanced Study and
Tufts university for their warm hospitalities during which a part of
this project was proceeded. J.~G. is supported by MEC
FPA2013-46570-C2-2-P, AGAUR 2014-SGR- 1474, and CPAN CSD2007-00042
Consolider-Ingenio 2010. Y.~U. is supported by JSPS Grant-in-Aid for Research
Activity Start-up under Contract No.~26887018 and Grant-in-Aid for
Scientific Research on Innovative Areas under Contract No.~16H01095. This research was
supported in part by Building of Consortia for the Development of Human Resources in
Science and Technology and the National Science Foundation under Grant No. NSF
PHY11-25915.

\appendix

\section{Consistency relations for $\zeta$ correlators} \label{Sec:ACR}
Using Eq.~(\ref{Cond:CR1}), we can derive the well-known consistency
relation for the curvature perturbation $\zeta$. Using
Eq.~(\ref{Cond:CR1}) for $n=3$ divided by the square of $\hat{W}^{(2)}(t; k_2)$, we
obtain 
\begin{align}
 & \frac{\hat{W}^{(3)}(t;\,0,\, k_2,\, k_2)}{\{\hat{W}^{(2)}(t;\, k_2)
 \}^2} - (\bm{k}_2 \cdot \partial_{\sbm{k}_2} + d)
 \frac{1}{\hat{W}^{(2)}(t,\, k_2)}=0\,. \label{Cond:CR13} 
\end{align}
Since the power spectrum and the bispectrum are given by
Eqs.~(\ref{Exp:Pzeta}) and (\ref{Exp:Bzeta}), respectively, we find that
Eq.~(\ref{Cond:CR13}) indeed gives the consistency relation for the bi-spectrum:
\begin{align}
 & \lim_{k_3/k_1,\, k_3/k_2 \to 0} \frac{B(t; k_1,\, k_2,\,
 k_3)}{P(k_3)} = - (\bm{k}_2 \cdot \partial_{\sbm{k}_2} + d) P(k_2) \,. 
\end{align}

\begin{figure}[t]
\begin{center}
\begin{tabular}{cc}
\includegraphics[width=16cm]{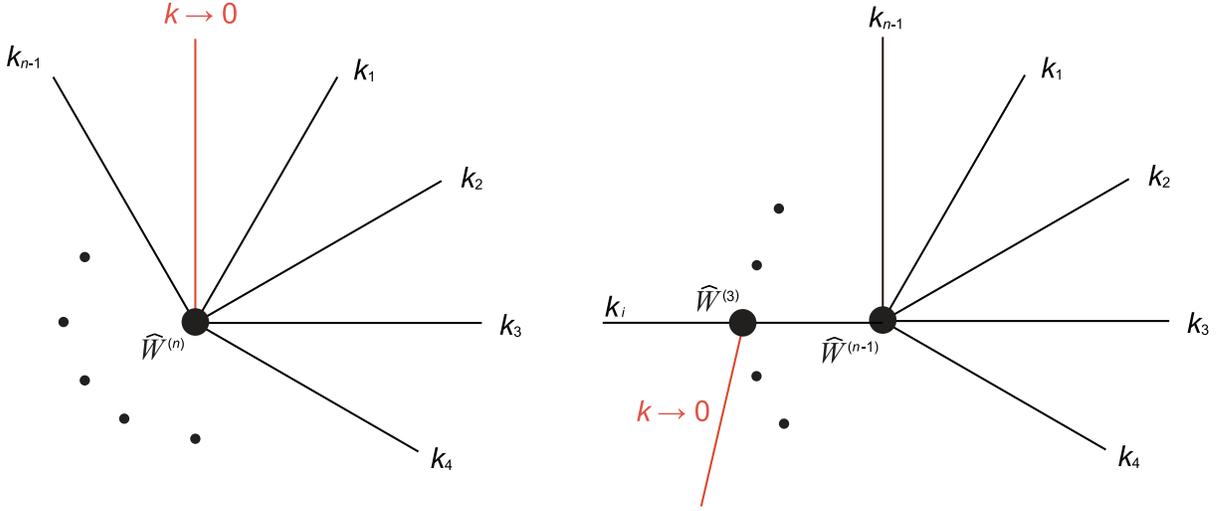}
\end{tabular}
\caption{The left diagram is the Feynman diagram for the first term in the first line of Eq. (A.4) and the right diagram is the one for the second one with the summation index $i$ .}
\label{Fg}
\end{center}
\end{figure}

A systematic derivation can be given by using the mathematical
induction. Dividing Eq.~(\ref{Cond:CR1}) by 
\begin{align}
 & \prod_{i=1}^{n-1} \hat{W}^{(2)}(k_i) = \hat{W}^{(2)} (K_{n-1}) \hat{W}^{(2)} (k_2) \cdots \hat{W}^{(2)}(k_{n-1})\,,
\end{align}
we obtain
\begin{align}
 & -\frac{\hat{W}^{(n)}(0,\, \{ \bm{k}_i \}_{n-1})}{ \prod_{i=1}^{n-1}
 \hat{W}^{(2)}(k_i)} + \left( \sum_{i=1}^{n-1} \bm{k}_i \cdot
 \partial_{\sbm{k}_i} \ln \frac{k_i^d}{\hat{W}^{(2)}(k_i)} \right)
 \frac{\hat{W}^{(n-1)}(\{ \bm{k}_i \}_{n-1})}{ \prod_{i=1}^{n-1} 
 \hat{W}^{(2)}(k_i)} \cr
 & = -  \left( \sum_{i=2}^{n-1} \bm{k}_i \cdot 
 \partial_{\sbm{k}_i} + d(n-2) \right) \left\{- \frac{\hat{W}^{(n-1)}(\{ \bm{k}_i \}_{n-1})}{ \prod_{i=1}^{n-1} 
 \hat{W}^{(2)}(k_i)}  \right\}  \,, \label{CompCR1}
\end{align}
where we used 
\begin{align}
 & \sum_{i=2}^{n-1} \bm{k}_i \cdot  \partial_{\sbm{k}_i}
 \hat{W}^{(2)}(K_{n-1}) = \bm{K}_{n-1} \cdot \partial_{\sbm{K}_{n-1}}
 \hat{W}^{(2)}(K_{n-1}) \,. 
\end{align}
The first term in the first line of Eq.~(\ref{CompCR1}) is the
contribution from the left diagram of Fig.~\ref{Fg} with one of the
$n$-momenta sent to 0. The minus sign appears, since the $n$-point
interaction is given by $- \hat{W}^{(n)}(\{ \bm{k}_i \}_n)$. Since the momentum derivative in the second term
can be rewritten as
\begin{align}
 &  \bm{k}_i \cdot \partial_{\sbm{k}_i} \ln \frac{k_i^d}{
 \hat{W}^{(2)}(k_i)}  = -f_{NL}^{local}(k_i) = \frac{\hat{W}^{(3)}(0,\, k_i,\,
 k_i)}{\hat{W}^{(2)}(k_i)}\,, 
\end{align}
we find that the second term is the contribution from the right diagram
of Fig.~\ref{Fg}. 

The expression in the braces in the second line denotes the contribution
from the one particle irreducible diagram for the $(n-1)$-point
function, which cannot be decomposed into two diagrams by cutting one of
the propagators included in the diagram. For $n=4$, Eq.~(\ref{CompCR1}) gives the consistency
relation (\ref{Exp:CRd}), since the three point function contains only the irreducible
diagram. For $n > 4$, the $(n-1)$-point function also contains reducible
diagrams and hence we obtain
\begin{align}
 & -\frac{\hat{W}^{(n)}(0,\, \{ \bm{k}_i \}_{n-1})}{ \prod_{i=1}^{n-1}
 \hat{W}^{(2)}(k_i)} + \left( \sum_{i=1}^{n-1} \bm{k}_i \cdot
 \partial_{\sbm{k}_i} \ln \frac{k_i^d}{\hat{W}^{(2)}(k_i)} \right)
 \frac{\hat{W}^{(n-1)}(\{ \bm{k}_i \}_{n-1})}{ \prod_{i=1}^{n-1} 
 \hat{W}^{(2)}(k_i)} \cr
 & \quad -  \left( \sum_{i=2}^{n-1} \bm{k}_i \cdot 
 \partial_{\sbm{k}_i} + d(n-2) \right) ( {\rm Contributions~from
 ~reducible~diagrams}) \cr 
 & = -  \left( \sum_{i=2}^{n-1} \bm{k}_i \cdot 
 \partial_{\sbm{k}_i} + d(n-2) \right)  {\cal C}^{(n-1)} (\{ \bm{k}_i \}_{n-1}) \,. \label{CompCR2}
\end{align}

\begin{figure}[t]
\begin{center}
\begin{tabular}{cc}
\includegraphics[width=11cm]{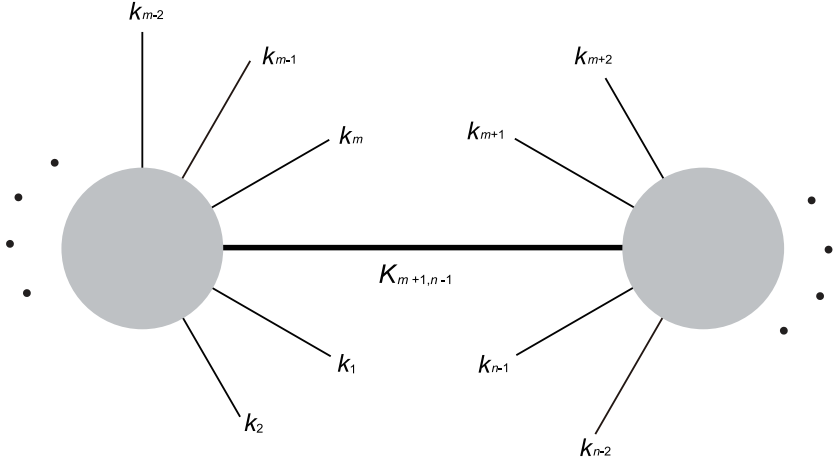}
\end{tabular}
\caption{A 1PI reducible diagram which shows up in the second line of Eq.~(A.7). We highlight the double counted propagator by the thick line.}
\label{Fg2}
\end{center}
\end{figure}

Now, all we need to show is that the summation of the first two
lines of Eq.~(\ref{CompCR2}) gives the left-hand side of the consistency
relation (\ref{Exp:CRd}). In order to compute the terms in the second
line, we consider a 1PI reducible diagram which is depicted in Fig.~\ref{Fg2}. This
diagram can be understood as connecting the two diagrams, the $(m+1)$-point function
with momenta $\{ {\bm k}_i \}_m$ and ${\bm K}_{m+1,\, n-1} =\sum_{i=m+1}^{n-1} \bm{k}_i$ and
$(n-m)$-point function with momenta $-{\bm K}_{m+1,\, n-1}$ and 
${\bm k}_j$ for $j=m+1,\, \cdots,\, n-1$. These two diagrams are jointed by the propagator with the
momentum ${\bm K}_{m+1,\, n-1}$. Here $m$ is $2 \leq m \leq n-3$. This contribution is given by
\begin{align}
 & {\cal C}^{(m+1)} ( {\bm K}_{2, n-1},\, \bm{k}_2,\, \cdots,\, \bm{k}_m,\,
 \bm{K}_{m+1,\, n-1}) \cr
 &\quad \times  {\cal C}^{(n-m)} ( \bm{K}_{m+1,\, n-1},\, \bm{k}_{m+1},\, \cdots,\, \bm{k}_{n-1})  \hat{W}^{(2)} (K_{m+1,\,
 n-1})\,, \label{reducible}
\end{align}
where the propagator $P(K_{m+1,\,n-1})= 1/\hat{W}^{(2)} (K_{m+1,\, n-1})$
is counted twice in the jointed two diagrams, so we divided by it. 
Operating $ \sum_{i=2}^{n-1} \bm{k}_i \cdot \partial_{\sbm{k}_i}$ on
Eq.~(\ref{reducible}) and using the consistency relation (\ref{Exp:CRd})
for the $(m+2)$-point and $(n-m+1)$-point functions in the squeezed limit,
we obtain
\begin{align}
 & - \left( \sum_{i=2}^{n-1} \bm{k}_i \cdot  \partial_{\sbm{k}_i} + d(n-2)
 \right) ({\rm Reducible~diagram~in~Fig.~1}) \cr
 & = \lim_{\sbm{k} \to 0} \frac{{\cal C}^{(m+2)}(\bm{k}, \{
 \bm{k}_i \}_m,\bm{K}_{m+1, n-1}) {\cal C}^{(n-m)} (
 \bm{K}_{m+1,\, n-1}, \bm{k}_{m+1}, \cdots, \bm{k}_{n-1})
 \hat{W}^{(2)} (K_{m+1,\, n-1}) }{P(k)}  \cr
 &\, +  \lim_{\sbm{k} \to 0}  \frac{{\cal C}^{(m+1)}(\{ \bm{k}_i
 \}_m,\bm{K}_{m+1, n-1}) {\cal C}^{(n-m+1)} ( \bm{k},
 \bm{K}_{m+1,\, n-1}, \bm{k}_{m+1}, \cdots, \bm{k}_{n-1}) \hat{W}^{(2)}
 (K_{m+1,\, n-1}) }{P(k)} \cr
 &\, - \lim_{\sbm{k} \to 0} \frac{1}{P(k)} \biggl[ {\cal C}^{(m+1)} (\{ \bm{k}_i
 \}_m, \bm{K}_{m+1,\, n-1}) \left( - \frac{\hat{W}^{(3)}(k, K_{m+1,\, n-1}, K_{m+1,\, n-1})}{\hat{W}^{(2)}(k)}  \right) \cr
 & \qquad \qquad \qquad \qquad \qquad \qquad \times  {\cal
 C}^{(n-m)} ( \bm{K}_{m+1,\, n-1},\, \bm{k}_{m+1},\, \cdots,\,
 \bm{k}_{n-1}  \biggr]\,.
\end{align} 
The term in the second line is the contribution from the diagram
with one soft leg inserted to the $(m+1)$-point function and the one
in the third line is from the diagram with one soft leg inserted to
the $(n-m)$-point function. Since the diagram with one soft leg inserted
to the joint propagator with the momentum $\bm{K}_{m+1,\, n-1}$ is
counted both in the second and third lines and hence the term in the
last two lines cancels the doubled counted one. In this way, we can see that the terms in
the first two lines of Eq.~(\ref{CompCR2}) add up the contributions form all the diagrams for 
${\cal C}^{(n)}(\bm{k},\, \{ \bm{k}_i\}_{n-1})$ in the squeezed limit,
$\bm{k} \to 0$. Now, the consistency relation for the $n$-point
function, which was derived in Ref.~\cite{Hinterbichler:2013dpa}, is
extended to an arbitrary spacetime dimension, following a different
method.


\end{document}